\documentclass[preprint,12pt]{elsarticle}

\usepackage{amssymb}
\usepackage{graphicx}
\usepackage{geometry} 
\usepackage{tabularx} 
\usepackage{booktabs} 
\usepackage{caption} 
\usepackage[font=small,labelfont=bf]{caption}
\usepackage[T1]{fontenc} 
\usepackage[utf8]{inputenc} 
\usepackage{amsmath} 
\usepackage{textcomp} 
\usepackage{mathptmx}
\usepackage{longtable}
\usepackage{bm}  
\usepackage{adjustbox} 
\usepackage{afterpage} 
\usepackage{lscape} 
\usepackage{graphicx}
\usepackage{subcaption}
\usepackage{hyperref}

\journal{Environment Research}

\begin{document}

\begin{frontmatter}



\title{Associations between exposure to OPEs and rheumatoid arthritis risk among adults in NHANES, 2011-2018}


\author[inst1]{Sneha Singh\footnote{sxs3309@mavs.uta.edu}}

\affiliation[inst1]{organization={Department of Earth and Environmental Sciences, The University of Texas at Arlington},
            addressline={701 S Nedderman Dr}, 
            city={Arlington},
            postcode={76019}, 
            state={TX},
            country={USA}}

\author[inst2]{Elsa Pirouz\footnote{elsa.pirouz@uta.edu}}
\author[inst3,inst4]{Amir Shahmoradi\footnote{a.shahmoradi@uta.edu}}

\affiliation[inst2]{organization={Department of Earth and Environmental Sciences, The University of Texas at Arlington},
            addressline={701 S Nedderman Dr}, 
            city={Arlington},
            postcode={76019}, 
            state={TX},
            country={USA}}
            
\affiliation[inst3]{organization={Department of Physics, The University of Texas at Arlington},
            addressline={701 S Nedderman Dr}, 
            city={Arlington},
            postcode={76019}, 
            state={TX},
            country={USA}}
\affiliation[inst4] {organization = {Division of Data Science, The University of Texas at Arlington}
                    , addressline={701 S Nedderman Dr}
                    , city={Arlington}
                    , postcode={76019} 
                    , state={TX}
                    , country={USA}
                    }

\begin{abstract}
    Rheumatoid arthritis (RA) has an intricate etiology that includes environmental factors as well as genetics. Organophosphate esters (OPEs) are frequently used as chemical additives in many personal care products and household items. However, there has been limited research on their potential effects on rheumatoid arthritis (RA). The specific associations between OPEs and RA remain largely unexplored. This study investigates any potential associations between adult rheumatoid arthritis risk and exposure to OPEs. We investigated data from the National Health and Nutrition Examination Survey (NHANES) 2011–2018 among participants over 20 years old. In two models, multivariable logistic regression was utilized to investigate the relationship between exposure to OPEs and RA. Furthermore, subgroup analyses stratified by age, gender, and dose exposure response were evaluated. Generalized additive models and smooth curve fits were used to characterize the nonlinear relationship between RA and OPEs.
    In conclusion, 5490 individuals (RA: 319, Non-RA: 5171) were analyzed. Higher quantiles (Q4) of DPHP and DBUP showed a higher prevalence of RA than the lowest quantiles. Our findings show that adult RA prevalence is higher in those who have been exposed to OPEs (DPHP, DBUP).
    Interestingly, these correlations seem to be stronger among women, the elderly, those with higher BMIs, and those who have diabetes. The dose-response curve for DPHP and DBUP demonstrated an upward-sloping trend. In contrast, BCEP and BCPP showed a U-shaped relationship and an inverted U-shaped relationship with the probability of RA.BDCPP demonstrated a complex relationship with a peak at lower concentrations followed by a decrease. Finally, our study also concludes that exposure to OPEs plays a crucial role in the pathogenesis of RA. 
\end{abstract}

\begin{graphicalabstract}
    \centering
    \includegraphics[width=1\textwidth]{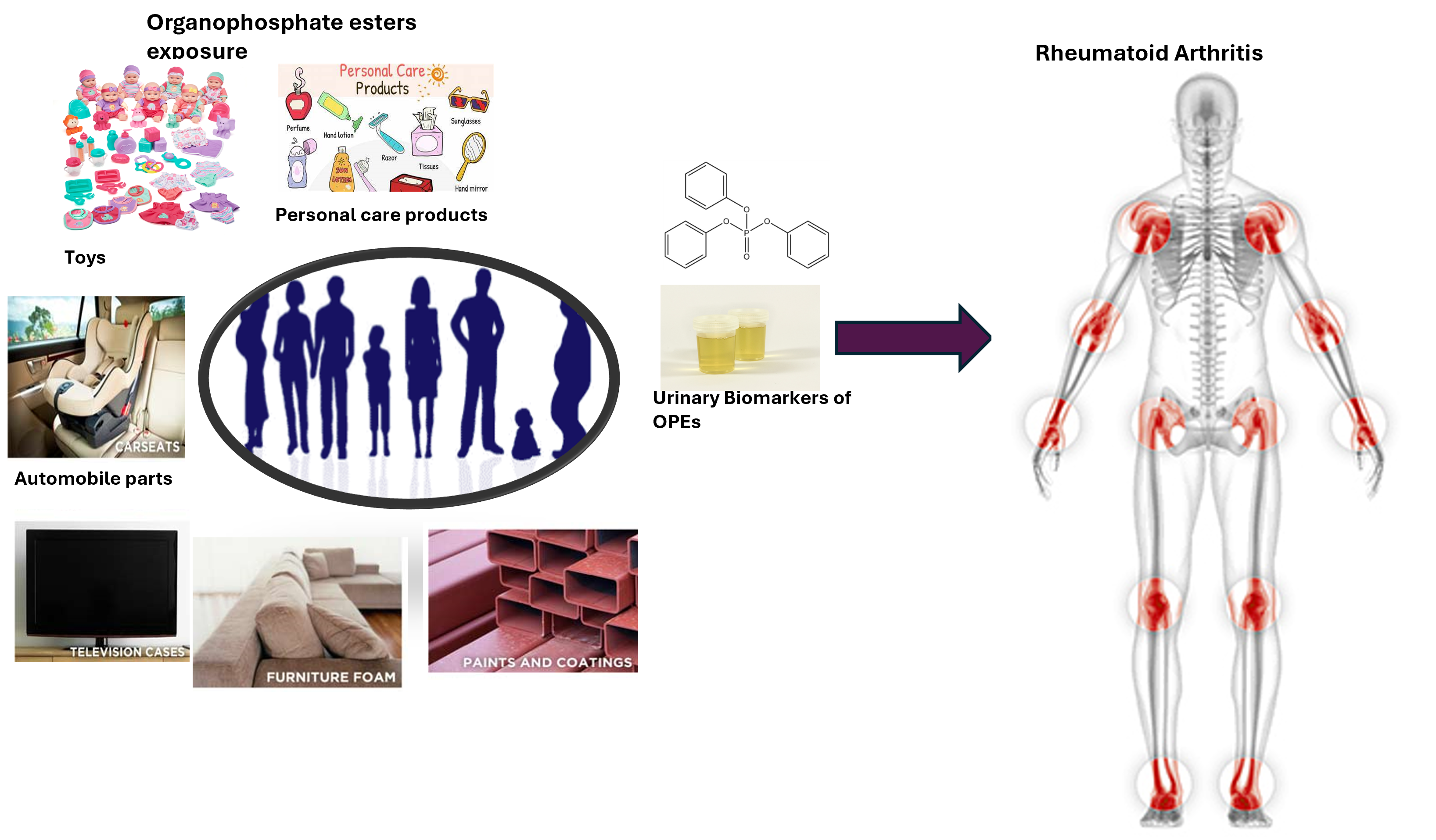}
\end{graphicalabstract}

\begin{highlights}
    \item
        Urinary metabolites of OPEs (DPHP, DBUP) are associated with the increased prevalence of RA.
    \item
        BDCPP appears to have a protective effect, reducing the risk of RA.
    \item
        Associations are more prominent in females, participants over 60 years, and those with higher BMI.
    \item
        Exposure to OPEs plays a critical role in the pathogenesis of rheumatoid arthritis (RA).
\end{highlights}


\end{frontmatter}


\section{Introduction}
\label{sec:sample1}
Rheumatoid arthritis (RA) is a chronic inflammatory autoimmune disorder that can cause irreversible joint damage and significant disability \cite{mori20111beta}. It is characterized by inflammatory changes of the synovial tissues of joints, cartilage, and bone \cite{scherer2020etiology}. Although the exact pathogenesis of RA is still unknown, it has become evident that the incidence and development of RA is a multi-factorial interaction and is closely associated with major genetic and environmental factors or stresses \cite{gabriel2001epidemiology}. According to previous studies, genetic and epigenetic influences accounted for approximately 50–60 \% of the risk for developing RA, while the remainder can be explained by environmental exposures such as dust exposure, smoking, and especially the microbiome, which also represents an "internal" environment \cite{de2018malignancy}. There appears to be an imperative reciprocation between the adaptive and innate immune system components. The association between environmental exposure and prevalence of RA is well established \cite{fang2023association, lei2023association, sun2020relationship, wang2023association}. Recent studies confirmed the augmenting evidence and substantiated that environmental pollutants like polycyclic aromatic hydrocarbons, phthalates, heavy metals, and volatile organic compounds play a quintessential role in the prevalence of RA\cite{chupeau2020organophosphorus}. Notably, exposures to pesticides, insecticides, and traffic-related pollutants have also been implicated in the heightened risk of RA, accentuating the intricate link between environmental pollution and autoimmune disease susceptibility \cite{gabriel2001epidemiology}. While RA is a lifelong condition, it can be controlled and prevented with some lifestyle modification, including regular exercise and a nutritious diet \cite{xu2021prevalence}.
Polybrominated diphenyl ethers (PBDEs) were banned in 2004 by the European Commission due to their environmental persistence, bio-accumulation, and associated toxicity. Organophosphate esters were introduced as an environmentally safer alternative to PBDEs \cite{blum2019organophosphate}. OPEs are organic phosphoric acid esters containing ether alkyl chains or aryl groups. They are either halogenated or nonhalogenated. Halogenated organophosphates are used as flame retardants, while nonhalogenated organophosphates are mostly used as plasticizers in consumer products, textiles, and construction materials \cite{van2012phosphorus}. OPEs are not chemically bound in the products they are added to, and due to this, they are released into the environment via volatilization, leaching, and/or abrasion \cite{blum2019organophosphate}. Human exposure to OPEs stems from various sources, including inhalation of dust, unintentional ingestion of contaminated foods, and dermal contact.
Indoor environments containing these chemicals serve as the most common source of exposure \cite{lei2023exposure}. Epidemiological studies revealed the adverse effects of OPEs on human health, such as endocrine disturbance, neurotoxicity, and carcinogenicity \cite{csaba2018effect}. Numerous adverse effects of exposure to OPEs have been identified in both experimental and human observational studies \cite{khani2023cellular}. Decreased total-body bone mineral density (BMD) and lumbar spine BMD in zebrafish and in vitro models, as well as exposure disrupted thyroid and sex hormones. Watanabe et al. showed that exposure to OPEs can increase the level of IFN-$\gamma$ in the bodily fluid of mice \cite{watanabe2017perinatal}. OPEs exposure activated interferon signaling pathways in rats that can affect at both at the level of biogenesis and bioactivity of immune ligands. Recent experimental studies provided significant insights into the linkage between OPEs exposure causing immunotoxicity and immunomodulatory effects on immune cells and humoral mediators \cite{dunnick2017tetrabromobisphenol}. The human body metabolizes OPEs rapidly and excretes them mainly through the urine, where diesters and monoesters are representative biomarkers \cite{bi2023organophosphate}. OPEs can damage the renal structure of mice through the NF-$\kappa$B p65/NLRP3-mediated inflammatory response \cite{tan2024relationship}. Although experimental studies demonstrate exposure to OPEs may alter the production of RA biomarkers cytokines IL-1$\beta$, IL-6, and TNF-$\alpha$ in macrophages. However, little is known about the relationship between OPEs exposure and RA in humans.

The present research conducts a cross-sectional study based on U.S. national population data from the National Health and Nutrition Examination Survey (NHANES), 2011 to 2018. The impact of urinary metabolites on OPEs was analyzed using multivariate logistic regression. Furthermore, weighted generalized additive models and restricted cubic splines were utilized to determine the non-linear relation between urinary metabolites and RA and identify the dose-response curve of each metabolite. Our results provide novel epidemiological evidence on the associations of urinary metabolites of OPEs exposures with RA risk and contribute to identifying the hazardous factors of RA.

\renewcommand{\textfraction}{0.1}  
\renewcommand{\topfraction}{0.85}  
\renewcommand{\bottomfraction}{0.7} 
\setcounter{topnumber}{3}          
\setcounter{bottomnumber}{2}       
\setcounter{totalnumber}{5}        

\section{Materials and Methods}

\subsection{Study design and population}
    
\begin{figure}[h!]
        \centering
        \includegraphics[width=1.1\textwidth]{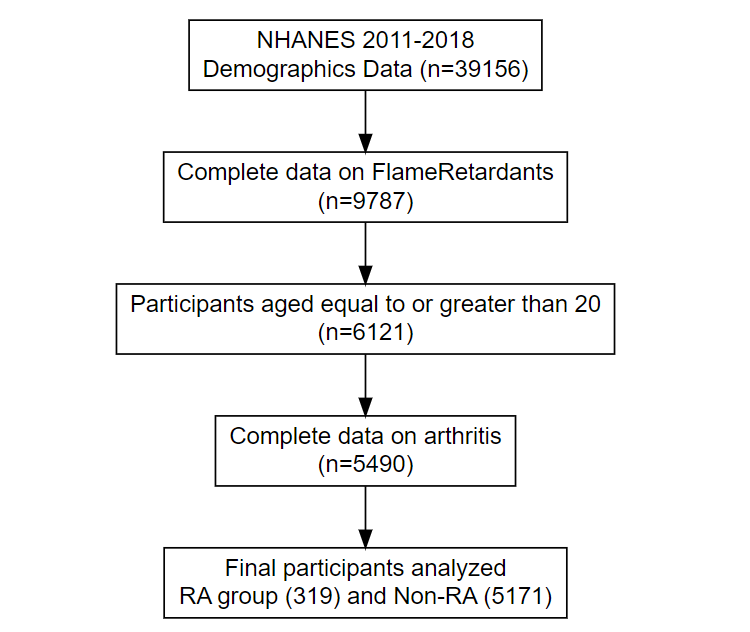}
        \caption{Flow chart of participant selection}
        \label{fig:flowchartRAFR}
    \end{figure}

    The National Health and Nutrition Examination Survey (NHANES) is a cross-sectional survey conducted by the National Centre for Health Statistics (NCHS) program of the Center for Disease Control (CDC), which is designed to assess the nutritional and health status of the non-institutionalized US population. A complex, stratified multistage sampling design is applied to make the sample more representative. The survey is conducted every two years, and information covering demographic, dietary, examination, laboratory, and questionnaire data was collected. In our study, urinary metabolites of OPEs were analyzed from survey cycles (2011-2012, 2013-2014, 2015-2016, 2017-2018). All the participants enrolled were provided informed consent at the time of recruitment for the survey. Detailed information about the participant's demographic, socio-economic, and dietary factors, as well as medical and health status, is collected through in-home interviews. A standardized medical and physical examination with bio-specimen (such as serum and urine) collection is conducted by highly trained medical personnel in specially equipped mobile examination centers (MECs). In our analysis, participants aged 20 years and older with complete data about urinary flame-retardant metabolites and rheumatoid arthritis were included for analysis. Finally, we have 5490 participants: 319 with rheumatoid arthritis and 5171 without rheumatoid arthritis (Figure \ref{fig:flowchartRAFR}).
    
    \begin{table}[!h]
        \centering
        \caption{Characteristics of the study population (N=5490), NHANES, 2011-2018.}
        \label{Table 1}
        \scriptsize 
        \begin{tabular}{llllll}
        \toprule
          Characteristics & Overall & Non-RA & RA & p & test \\
        \midrule
        n & 5490 & 5171 & 319 &  & \\
        Gender = Female (\%) & 2793 (50.9) & 2603 (50.3) & 190 (59.6) & \textbf{0.002} & \\
        Age (mean (SD)) & 48.35 (17.47) & 47.60 (17.42) & 60.51 (13.27) & \textbf{<0.001} & \\
        Ethnicity (\%) &  &  &  & \textbf{<0.001} & \\
        Other race & 779 (14.2) & 730 (14.1) & 49 (15.4) &  & \\
        \addlinespace
        Mexican American & 754 (13.7) & 705 (13.6) & 49 (15.4) &  & \\
        Non-Hispanic White & 2083 (37.9) & 1972 (38.1) & 111 (34.8) &  & \\
        Non-Hispanic Asian & 715 (13.0) & 700 (13.5) & 15 (4.7) &  & \\
        Non-Hispanic Black & 1159 (21.1) & 1064 (20.6) & 95 (29.8) &  & \\
        Marital\_status (\%) &  &  &  & \textbf{<0.001} & \\
        \addlinespace
        Married & 3285 (59.8) & 3116 (60.3) & 169 (53.0) &  & \\
        Missing & 1 (0.0) & 1 (0.0) & 0 (0.0) &  & \\
        Never married & 1095 (19.9) & 1060 (20.5) & 35 (11.0) &  & \\
        Previously married & 1109 (20.2) & 994 (19.2) & 115 (36.1) &  & \\
        Education (\%) &  &  &  & \textbf{<0.001} & \\
        \addlinespace
        Above high school & 3134 (57.1) & 2989 (57.8) & 145 (45.5) &  & \\
        High school or equivalent & 1225 (22.3) & 1145 (22.1) & 80 (25.1) &  & \\
        Missing & 7 (0.1) & 6 (0.1) & 1 (0.3) &  & \\
        Under high school & 1124 (20.5) & 1031 (19.9) & 93 (29.2) &  & \\
        Citizenship (\%) &  &  &  & \textbf{0.002} & \\
        \addlinespace
        Citizen & 3772 (68.7) & 3524 (68.1) & 248 (77.7) &  & \\
        Missing & 1 (0.0) & 1 (0.0) & 0 (0.0) &  & \\
        Not a citizen & 1717 (31.3) & 1646 (31.8) & 71 (22.3) &  & \\
        \addlinespace
        Urinary\_creatinine (mean (SD)) & 123.34 (81.41) & 123.50 (80.98) & 120.67 (88.15) & 0.546 & \\
        Alcohol (\%) &  &  &  & \textbf{0.008} & \\
        \addlinespace
        Missing & 422 (7.7) & 404 (7.8) & 18 (5.6) &  & \\
        No & 1186 (21.6) & 1096 (21.2) & 90 (28.2) &  & \\
        Yes & 3882 (70.7) & 3671 (71.0) & 211 (66.1) &  & \\
        BMI (mean (SD)) & 29.20 (7.12) & 29.05 (7.07) & 31.69 (7.43) & \textbf{<0.001} & \\
        Diabetes (\%) &  &  &  & \textbf{<0.001} & \\
        \addlinespace
        Missing & 5 (0.1) & 3 (0.1) & 2 (0.6) &  & \\
        No & 4693 (85.5) & 4475 (86.5) & 218 (68.3) &  & \\
        Yes & 792 (14.4) & 693 (13.4) & 99 (31.0) &  & \\
        Smoking (\%) &  &  &  & \textbf{0.001} & \\
        Every day & 858 (15.6) & 801 (15.5) & 57 (17.9) &  & \\
        \addlinespace
        Missing & 3196 (58.2) & 3043 (58.8) & 153 (48.0) &  & \\
        Not at all & 1207 (22.0) & 1111 (21.5) & 96 (30.1) &  & \\
        Some days & 229 (4.2) & 216 (4.2) & 13 (4.1) &  & \\
        Age.cat = greaterthan 60 years (\%) & 1670 (30.4) & 1489 (28.8) & 181 (56.7) & \textbf{<0.001} & \\
        Activity\_level (\%) &  &  &  & \textbf{0.011} & \\
        \addlinespace
        Moderate Activity & 565 (10.3) & 540 (10.4) & 25 (7.8) &  & \\
        None & 4641 (84.5) & 4354 (84.2) & 287 (90.0) &  & \\
        Vigorous Activity & 284 (5.2) & 277 (5.4) & 7 (2.2) &  & \\
        PIR (\%) &  &  &  & \textbf{<0.001} & \\
        1-1.99 & 1315 (24.0) & 1239 (24.0) & 76 (23.8) &  & \\
        \addlinespace
        2-3.99 & 1284 (23.4) & 1220 (23.6) & 64 (20.1) &  & \\
        Greater than or equal to 4 & 1314 (23.9) & 1264 (24.4) & 50 (15.7) &  & \\
        Less than 1 & 1073 (19.5) & 973 (18.8) & 100 (31.3) &  & \\
        Missing & 504 (9.2) & 475 (9.2) & 29 (9.1) &  & \\
        \bottomrule
        \end{tabular}
        \caption*{Note: BMI, body mass index; PIR, Poverty income ratio; SD, standard deviation; P-value in bold indicates <0.05.}
    \end{table}
    
\subsection{Assessment of RA}
    
    The diagnosis of RA status was assessed by self-report in a face-to-face interview during a questionnaire examination regarding the health condition. Arthritis-related questions in the medication condition questionnaire were completed before a physical examination using the computer-assisted personal interviewing (CAPI) system. At first, participants were asked, “Has a doctor or other health professional ever told you that you have arthritis?” stored under MCQ160A, and If the answer was “yes,” a follow-up question was asked: “Which type of arthritis was it? “Stored as MCQ195. Participants were classified as RA or Non-RA according to their answers to the second question.
    
\subsection{Detection of OPEs and creatinine in urine}
    
    The urine specimens of enrolled participants were collected, aliquoted, and shipped to the National Center for Environmental Health (NCEH) according to the analytical guidelines. Due to the target analytes being diverse in different NHANES cycles, we selected five consecutively detected flame retardants for statistical analysis: diphenyl phosphate (DPhP), dibutyl phosphate (DBuP), bis(1,3-dichloro-2-propyl) phosphate (BDCIPP), bis(1-chloro-2-propyl) phosphate (BCIPP), bis(2-chloroethyl) phosphate (BCEP), and tetrabromobisphenol A (TBBA). Briefly, the analytical method is based on enzymatic hydrolysis of urinary conjugates of target analytes, automated offline solid phase extraction, and isotope dilution high-performance liquid chromatography-tandem mass spectrometry detection. The lower detection limits (LLOD) for urinary BCIPP, BDCPP, DBuP, and DPhP were 0.1 µg/L, and the LLOD for urinary TBBA was 0.05 µg/L.The creatinine in the urine is determined using an Enzymatic Roche coba 6000 analyzer (Roche Di-agnostics, Indianapolis, IN, USA). Urinary creatinine concentrations were measured to account for dilution-dependent sample variation in biomarker concentrations.\textbf{Table}~\ref{Table 2} In our research, we selected chemicals for inclusion based on the metabolite’s detectable frequencies being greater than 50\%. The distributions of flame retardant metabolite concentrations were described using detection frequency, GM (95\% CI), and selected percentiles.
    
    \afterpage{
        \clearpage
        \begin{landscape} 
            \begin{table}[htbp]
                \small 
                \centering
                \caption{Distribution of urinary OPEs metabolites (µg/L) in US adults, NHANES 2011-2018.}
                \label{Table 2}
                \begin{tabular}{lcccccccc}
                \toprule
                \textbf{OPEs (µg/L)} & \textbf{Detection frequency (\%)} & \textbf{GM (95\% CI)} & \textbf{Min} & \textbf{25th} & \textbf{50th} & \textbf{75th} & \textbf{Max} \\
                \midrule
                DPHP & 95.88 & 0.74 (0.72, 0.77) & 0.070 & 0.33 & 0.74 & 1.57 & 126 \\
                BDCPP & 93.75 & 0.81 (0.79, 0.84) & 0.070 & 0.34 & 0.82 & 1.95 & 260 \\
                BCPP & 51.93 & 0.15 (0.15, 0.16) & 0.070 & 0.070 & 0.11 & 0.29 & 26.1 \\
                BCEP & 82.17 & 0.38 (0.37, 0.40) & 0.070 & 0.154 & 0.36 & 0.871 & 156 \\
                DBuP & 51.07 & 0.13 (0.13, 0.13) & 0.070 & 0.070 & 0.070 & 0.23 & 24.9 \\
                \bottomrule
                \end{tabular}
                \caption*{GM: geometric mean; DF: detection frequency; \\ Diphenyl phosphate (DPHP); Bis(1,3-dichloro-2-propyl) phosphate (BDCPP);\\ Bis(1-choloro-2-propyl) phosphate (BCPP); Bis(2-chloroethyl) phosphate (BCEP);\\ Dibutyl phosphate (DBUP).}
            \end{table}
        \end{landscape}
        \clearpage 
    }

\subsection{Assessment of Covariates}

    Directed Acyclic Graphs (DAGs) are developed based on existing literature and assumptions regarding plausible causal relationships among the covariates. We used the Daggity web application to establish the minimally sufficient adjustment sets of variables to estimate the association and total effects of OPEs on Rheumatoid Arthritis. Covariates included Age( 20-60 years, Greater than 60 years), Gender(Female and Male), Ethnicity(Mexican American, Non- Hispanic White, Non- Hispanic Black, Non- Hispanic Asian, and other race), Education(Under high school, Above high school, High school or equivalent, ), Marital status(Previously married, Married, and Never Married), the ratio of family income to poverty (PIR) was binned into four levels(Less than 1, 1-1.99, 2-3.99, Greater than or equal to 4), Alcohol status(Yes, No), Smoking(Every day, Some days, Not at all), body mass index (BMI) (kg/m$^2$), Urinary Creatinine(mg/dL), Activity level(Vigorous Activity, Moderate Activity, None), Citizenship(Citizen, Non- citizen), Diabetes(Yes, No). Missing data for the covariates were coded as a missing indicator category for categorical variables (citizenship (n = 1), education levels (n = 7), marital status (n = 1), smoking (n = 3196), and alcohol status (n = 422)) and were imputed with the median for the continuous variables (BMI (n = 48), family income (n = 504) and urinary creatinine (n = 1).

\subsection{Statistical Analysis}

    Continuous variables in covariates were described as means ± standard deviation and analyzed using the Mann-Whitney U test. Univariate analyses were used to assess exposure and covariate distributions. Categorical variables were expressed as numbers (percent) and compared by the Chi-square test (X\(^2\)) or Fisher exact test. All OPEs exposure variables were natural-log-transformed due to a long right tail before running a multivariate logistics regression. Urinary OPEs metabolites were stratified based on quantiles (Q1: $<$ 25th percentile, Q2: $\geq$ 25th to 50th percentile, Q3: $\geq$ 50th to 75th percentile, Q4: $\geq$ 75th percentile). For multivariate logistic regression, we used sampling weights to account for the complex NHANES survey design. Firstly, we fitted a crude model that assessed the association between each urinary metabolite of OPEs and RA status by comparing the second, third, and fourth quantiles to the first metabolite concentration using multivariate logistic regression. After that, two separate models were conducted for logistic regression: model 1 was adjusted for age and gender, and model 2 was adjusted for ethnicity, education, marital status, citizenship, PIR, alcohol status, smoking, activity level, BMI, diabetes, and urinary creatinine. Stratification analysis was also performed to evaluate the relationship between RA and OPEs in the US population by subgrouping the age and gender respectively based on the fact that incidence of RA majorly varies in different age group and gender.
    Additionally, Pearson correlations were performed on urinary OPEs to calculate the degree of correlation among them. We also assessed the nonlinear relationship between RA and OPEs with smooth curve fittings using generalized additive models (GAM). GAM is a flexible and smooth technique that captures the non-linearities in the data and helps us to fit nonlinear models. We implemented GAMs in R using the “gam” package. They are generalized versions of the linear model in which the predictors \(X_i\) depend either linearly or non-linearly on smooth non-linear functions like splines, polynomials, or step functions.
    Mathematical representation for Generalized additive models the regression equation:
    \begin{equation}
        \text f(x) = y_i = \alpha + f_{1}(x_{i1}) + f_{2}(x_{i2}) + \cdots + f_{p}(x_{ip}) + \epsilon_i 
    \end{equation}
    where the functions \( f_{1}, f_{2}, f_{3} \ldots f_{p} \) are different non-linear functions on variables \( X_{p} \). Variance inflation factors (VIF) and Spearman’s correlation analysis were employed to examine collinearity among variables for each model, respectively. A VIF value below 5 was considered indicative of the absence of multicollinearity \cite{berry2020human}. Although GAM models are capable of offering a flexible and detailed exploratory analysis of the data by helping to identify complex and nonlinear relationships between OPEs exposures and RA risk without mandating strict assumptions about the shape of the dose-response curve. Restricted cubic spline (RCS) allows for precise modeling of the dose-response relationship, providing a detailed visualization of how variations in OPEs exposure levels correlate with changes in RA risk. This approach facilitates the identification of possible thresholds and critical exposure levels. We selected three knots for each OPE as the default and used the Wald test to identify nonlinear dose-response relationships. The logistic model with restricted cubic splines can be written as:
    \begin{equation}
        \label{eq:logit}
        \text{logit}(P(D = 1 \mid V)) = \beta_0 + \beta_1 V + \beta_2 h_1(V) + \beta_3 h_2(V)
    \end{equation}
    Where \( h_1(V) \) and \( h_2(V) \) are the spline basis functions defined by the chosen knots.
    Adjusted factors were the same as fully adjusted models. RCS analysis was performed by \texttt{rms} R package (version 6.7)

\section{Results}

\subsection{Characteristics of the study population}

    This study included a total of 5490 adults, distinguished on the basis of RA status. Participants with rheumatoid arthritis (RA) (n = 319) and those without RA (Non-RA) (n = 5171). The mean age of participants was 48.35 years, with 50.9\% being female. Participants with RA were of older age, predominantly female, non-Hispanic Black, and with higher BMI as compared to non-RA. We observed a higher prevalence of diabetes and less physical activity (\textit{P} $<$ 0.05). No significant differences were observed in urinary creatinine and alcohol consumption (\textit{P} $>$ 0.05).

\subsection{Measurement of urinary metabolites of OPEs}

    The detection frequency, mean concentration, geometric mean concentration, and distribution of OPES are shown in Table 2.TBBA was excluded in subsequent analyses because 99.78\% samples were below the LLOD. The detection frequency of the remaining flame retardants was above 50\%, including DPHP (95.88\%), BDCPP (93.75\%), BCEP (82.17\%), DBUP (51.07\%) and BCPP (51.93\%). OPEs exposure patterns varied largely across population subgroups. Higher levels of DPHP and DBUP were observed in females (P$<$0.05), whereas BDCPP and BCEP levels were higher in males (P$<$0.05). We observed a significant positive correlation among each pair of OPEs (P$<$ 0.001), with the highest correlation between DPHP and BDCPP (coefficients = 0.53).

    \begin{figure}[h!]
        \centering
        \includegraphics[width=1\linewidth]{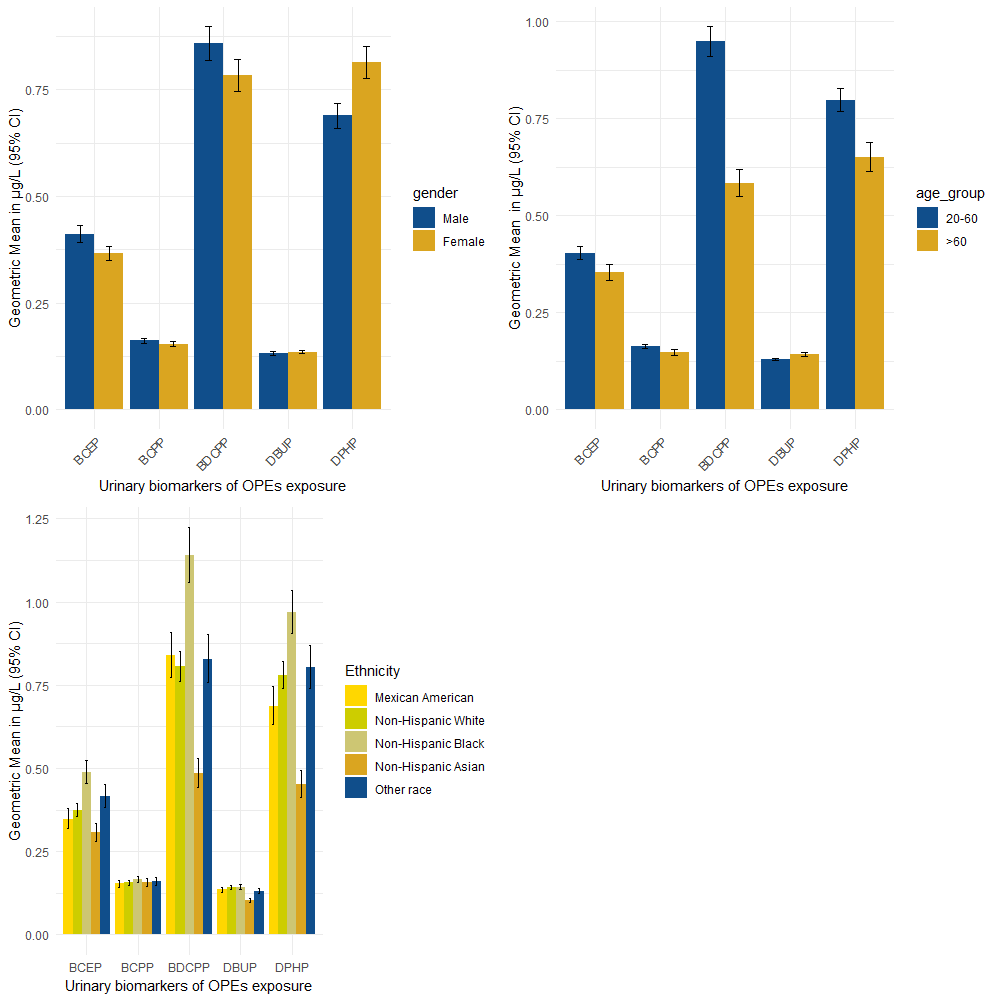}
        \caption{Urinary OPEs concentrations by demographic groups: gender (top-left), age (top-right), ethnicity (bottom). The bar plots represent the geometric means of each OPEs and the associated 95\% 
        confidence intervals.\label{fig:barplot1000by1000}}
     \end{figure}
    
\subsection{Association between urinary metabolites of OPEs and RA}
    
    Based on a sample size of 5,490 observations, the correlation analysis of five urinary metabolites, DPHP, BDCPP, BCPP, BCEP, and DBUP, revealed multiple significant relationships. BDCPP (r = 0.25), BCPP (r = 0.13), BCEP (r = 0.16), and DBUP (r = 0.10) showed positive associations with DPHP. Positive correlations were found between BDCPP and BCPP (r = 0.15), BCEP (r = 0.20), and DBUP (r = 0.02). The BCPP and BCEP showed the strongest positive association (r = 0.28), indicating a moderate link. Except for the correlation between DBUP and BDCPP, which was not significant after correction (p = 0.07), all correlations were statistically significant (p < 0.05). These results suggest that there may be shared sources or comparable metabolic pathways between higher levels of one metabolite and higher levels of another. In model 1, adjusted for age and gender, the relationship between different quartiles of the flame-retardant metabolite DPHP and the risk of RA was examined. Compared to Q1, participants with the increasing quantile of DPHP (Q3, OR: 1.56, 95\% CI: 0.95-2.56), DBUP (Q2, OR: 1.52, 95\% CI: 1.03-2.24; Q3, OR: 1.45, 95\% CI: 0.96-2.21; Q4, OR: 1.60, 95\% CI: 1.00-2.57) showed significantly higher risk of RA compared to Q1 and BDCPP (Q3, OR: 0.47, 95\% CI: 0.29-0.76) showed lower risk of RA as compared to Q1. In model 2, which was adjusted for age category, gender, BMI, ethnicity, citizenship, education, marital status, Poverty income ratio, smoking, diabetes, alcohol, activity level, and urinary creatinine (log-transformed), BDCPP showed a lower risk of RA compared to (Q4, OR: 0.55, 95\% CI: 0.33-0.91) (all p for trend<0.05) compared to Q1. 
    Our methodology's simultaneous use of GAMs and RCS ensures an extensive examination of the association between OPEs exposure and RA. The initial exploratory observations were provided by GAMs, which highlighted plausible nonlinear patterns. The dose-response relationship was investigated in detail and with specificity by RCS, significantly strengthened the findings' accuracy and interoperability.
    \\

    We utilized the generalized additive model (GAM) to explore the nonlinear relationship between RA and OPEs. The smooth function for BDCPP shows a slight non-linear pattern with a p-value of 0.711 estimated edf; 1.5, with a minor upward trend at higher levels. The smooth function for BCEP shows a notable non-linear pattern, with an initial decrease followed by an increase having a p-value of 0.53 and edf; 2.44. Variance inflation factors were calculated for gam utilizing \texttt{vif.gam()} function from \texttt{mgcv.helper} package to examine collinearity among variables. The plot from the RCS revealed distinct characteristics between OPEs and RA. DPHP and DBUP showed a positive association with increasing log odds of RA at higher concentrations. In contrast, BCEP showed a U-shaped relationship, suggesting a higher incidence of RA at both lower and higher concentrations with a huge dip in the center.

\begin{table}[htbp]
    \centering
    \caption{ORs (95\% CI) from Multivariable logistic regression analysis models of the association between OPEs and risk of RA.}
    \scriptsize 
    \label{tab:models}
    \begin{adjustbox}{width=\textwidth}
    \begin{tabular}{lcccccc}
    \toprule
    \textbf{OPEs}&  \multicolumn{2}{c}{\textbf{Model 1$^a$}} & \multicolumn{2}{c}{\textbf{Model 2$^b$}} \\
     & \textbf{OR (95\% CI)} & \textit{P-value} & \textbf{OR (95\% CI)} & \textit{P-value} \\
    \midrule
    DPHP & & & & \\
    Quantile 1 & 1.00 & & 1.00 & \\
    Quantile 2 & 1.23 (0.73; 2.09) & 0.441 & 1.21 (0.72; 2.06) & 0.474 \\
    Quantile 3 & 1.56 (0.95; 2.56) & 0.083 & 1.48 (0.92; 2.37) & 0.114 \\
    Quantile 4 & 1.33 (0.81; 2.17) & 0.263 & 1.22 (0.77; 1.93) & 0.402 \\
    \midrule
    BDCPP & & & & \\
    Quantile 1 & 1.00 & & 1.00 & \\
    Quantile 2 & 1.04 (0.67; 1.62) & 0.859 & 1.13 (0.72; 1.78) & 0.590 \\
    Quantile 3 & 0.79 (0.50; 1.27) & 0.336 & 0.91 (0.55; 1.48) & 0.691 \\
    Quantile 4 & 0.47 (0.29; 0.76) & \textbf{0.003} & 0.55 (0.33; 0.91) & \textbf{0.024} \\
    \midrule
    BCPP & & & & \\
    Quantile 1 & 1.00 & & 1.00 & \\
    Quantile 2 & 0.87 (0.56; 1.36) & 0.541 & 0.86 (0.54; 1.37) & 0.528 \\
    Quantile 3 & 0.89 (0.54; 1.48) & 0.659 & 0.88 (0.52; 1.49) & 0.642 \\
    Quantile 4 & 1.31 (0.75; 2.31) & 0.348 & 1.30 (0.73; 2.29) & 0.377 \\
    \midrule
    BCEP & & & & \\
    Quantile 1 & 1.00 & & 1.00 & \\
    Quantile 2 & 0.83 (0.52; 1.31) & 0.427 & 0.85 (0.53; 1.35) & 0.498 \\
    Quantile 3 & 1.16 (0.75; 1.81) & 0.513 & 1.13 (0.72; 1.78) & 0.597 \\
    Quantile 4 & 1.12 (0.70; 1.80) & 0.643 & 1.11 (0.69; 1.80) & 0.670 \\
    \midrule
    DBUP & & & & \\
    Quantile 1 & 1.00 & & 1.00 & \\
    Quantile 2 & 1.52 (1.03; 2.24) & \textbf{0.039} & 1.35 (0.92; 1.98) & 0.133 \\
    Quantile 3 & 1.45 (0.96; 2.21) & \textbf{0.085} & 1.25 (0.81; 1.93) & 0.328 \\
    Quantile 4 & 1.60 (1.00; 2.57) & \textbf{0.055} & 1.29 (0.78; 2.13) & 0.325 \\
    \bottomrule
    \end{tabular}
    \end{adjustbox}
    \caption*{A Logistic regression adjusted for \texttt{age.cat} and gender.\\
    b Logistic regression adjusted for \texttt{age.cat}, gender, BMI, ethnicity, citizenship, education, marital status, Poverty income ratio, smoking, diabetes, alcohol, activity level, urinary creatinine (log-transformed).}
\end{table}

\subsubsection{Subgroup analysis}

    The covariates adjusted in the subgroup analysis were \texttt{age.cat}, gender, BMI, ethnicity, citizenship, education, marital status, poverty income ratio, smoking, diabetes, alcohol, activity level, urinary creatinine (log-transformed), and the results from the subgroup analysis showed the association between flame retardants quantiles and RA was mainly present in females and in participants aged more than 60 years. Female had a growing risk of RA in the population with increasing quantiles of DPHP (Q2, OR: 2.52, 95\% CI: 1.32-4.81; Q3, OR: 3.07, 95\% CI: 1.37-6.88; Q4, OR: 4.04, 95\% CI: 1.42-11.46), BCEP (Q3, OR: 2.67, 95\% CI: 1.03-6.95), DBUP (Q4, OR: 2.92, 95\% CI: 0.94-9.07) compared to Q1. An increased risk of RA was observed in participants aged 20-60 years and greater than 60 years. For DPHP (Q4, OR: 3.50, 95\% CI: 0.96-12.70), BCPP (Q2, OR: 2.61, 95\% CI: 1.02-6.68; Q3, OR: 3.48, 95\% CI: 0.96-12.67; Q4, OR: 11.48, 95\% CI: 1.38-5.69), DBUP (Q4, OR: 5.89, 95\% CI: 1.03-3.68). The growing risk of RA in males was insignificant for most metabolites and participants over 60 years.

    \begin{figure}[h!]
        \centering
        \includegraphics[width=1\linewidth]{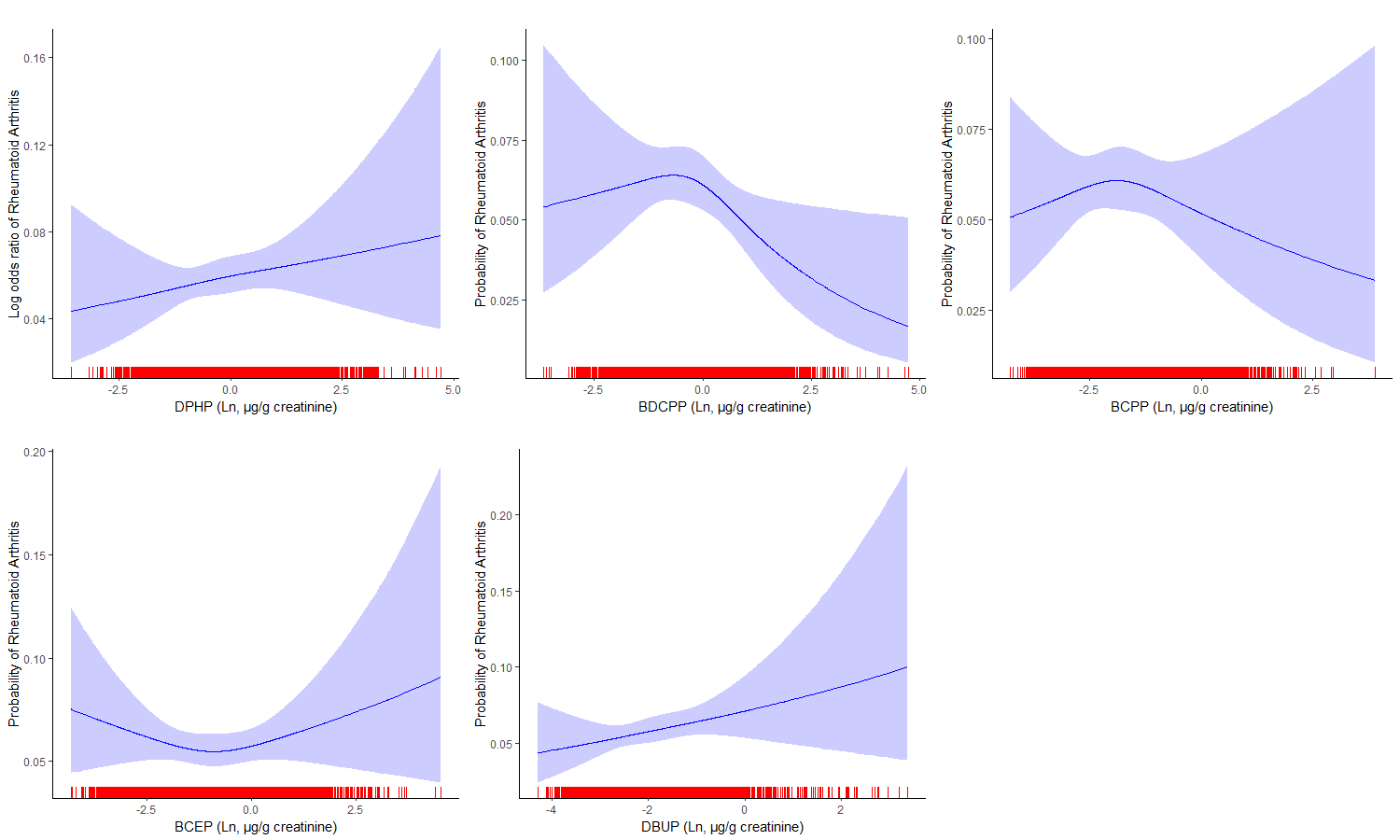}
        \caption{Dose-response relationships between urinary OPEs metabolites concentrations and Rheumatoid Arthritis. Adjusted for \texttt{age.cat}, gender, BMI, ethnicity, citizenship, education, marital status, poverty income ratio, smoking, diabetes, alcohol, activity level, and urinary creatinine (log-transformed). The red rugs represent the distribution of concentration.}
         \label{fig:doseResponseCurve}
     \end{figure}

    \begin{table}[htbp]
        \centering
        \scriptsize 
        \caption{ORs (95\% CI) from subgroup analysis between RA and OPEs stratified by gender.}
        \label{tab:gender_subgroup}
        \begin{adjustbox}{width=\textwidth}
        \begin{tabular}{lcccccc}
        \toprule
        \textbf{OPEs} & \multicolumn{2}{c}{\textbf{Male (n= 2697)}} & \multicolumn{2}{c}{\textbf{Female (n= 2793)}} \\
         & \textbf{OR (95\% CI)} & \textbf{P value} & \textbf{OR (95\% CI)} & \textbf{P value} \\
        \midrule
        DPHP & & & & \\
        Quantile 1 & 1.00 & & 1.00 & \\
        Quantile 2 & 0.90 (0.32; 2.56) & 0.849 & 2.52 (1.32; 4.81) & 0.007 \\
        Quantile 3 & 1.19 (0.30; 4.79) & 0.806 & 3.07 (1.37; 6.88) & 0.009 \\
        Quantile 4 & 0.99 (0.10; 9.35) & 0.990 & 4.04 (1.42; 11.46) & 0.011 \\
        \midrule
        BDCPP & & & & \\
        Quantile 1 & 1.00 & & 1.00 & \\
        Quantile 2 & 1.22 (0.46; 3.27) & 0.690 & 1.07 (0.51; 2.26) & 0.857 \\
        Quantile 3 & 1.23 (0.30; 5.15) & 0.773 & 0.72 (0.25; 2.08) & 0.545 \\
        Quantile 4 & 0.52 (0.05; 5.77) & 0.593 & 0.53 (0.12; 2.25) & 0.394 \\
        \midrule
        BCPP & & & & \\
        Quantile 1 & 1.00 & & 1.00 & \\
        Quantile 2 & 1.41 (0.58; 3.41) & 0.450 & 1.25 (0.62; 2.49) & 0.535 \\
        Quantile 3 & 1.42 (0.32; 6.28) & 0.644 & 1.65 (0.79; 3.42) & 0.187 \\
        Quantile 4 & 5.98 (0.70; 51.28) & 0.109 & 2.46 (0.78; 7.80) & 0.133 \\
        \midrule
        BCEP & & & & \\
        Quantile 1 & 1.00 & & 1.00 & \\
        Quantile 2 & 1.29 (0.61; 2.71) & 0.510 & 1.07 (0.45; 2.54) & 0.872 \\
        Quantile 3 & 2.67 (1.03; 6.95) & 0.049 & 1.44 (0.58; 3.60) & 0.436 \\
        Quantile 4 & 2.32 (0.67; 8.06) & 0.191 & 2.08 (0.48; 9.04) & 0.334 \\
        \midrule
        DBuP & & & & \\
        Quantile 1 & 1.00 & & 1.00 & \\
        Quantile 2 & 1.57 (0.67; 3.67) & 0.307 & 1.47 (0.75; 2.87) & 0.268 \\
        Quantile 3 & 1.78 (0.51; 6.20) & 0.372 & 1.60 (0.73; 3.48) & 0.244 \\
        Quantile 4 & 1.70 (0.19; 15.47) & 0.638 & 2.92 (0.94; 9.07) & 0.070 \\
        \bottomrule
        \end{tabular}
        \end{adjustbox}
        \caption*{Gender Subgroup Model: Logistic regression adjusted for \texttt{age.cat}, gender, BMI, ethnicity, citizenship, education, marital status, Poverty income ratio, smoking, diabetes, alcohol, activity level, and urinary creatinine (log-transformed)}
    \end{table}

    \begin{table}[htbp]
        \centering
        \scriptsize 
        \caption{OPEs for Age Subgroups (20-60 years and >60 years)}
        \label{tab:age_subgroup}
        \begin{adjustbox}{width=\textwidth}
        \begin{tabular}{lcccccc}
        \toprule
        \textbf{OPEs} & \multicolumn{2}{c}{\textbf{Age (20-60 years) (n= 3820)}} & \multicolumn{2}{c}{\textbf{Age (>60 years) (n= 1670)}} \\
         & \textbf{OR (95\% CI)} & \textbf{P value} & \textbf{OR (95\% CI)} & \textbf{P value} \\
        \midrule
        DPHP & & & & \\
        Quantile 1 & 1.00 & & 1.00 & \\
        Quantile 2 & 1.55 (0.61; 3.97) & 0.363 & 1.60 (0.83; 3.08) & 0.168 \\
        Quantile 3 & 2.45 (0.86; 6.97) & 0.099 & 1.64 (0.65; 4.13) & 0.302 \\
        Quantile 4 & 1.84 (0.43; 7.96) & 0.417 & 3.50 (0.96; 12.70) & 0.062 \\
        \midrule
        BDCPP & & & & \\
        Quantile 1 & 1.00 & & 1.00 & \\
        Quantile 2 & 1.14 (0.45; 2.84) & 0.787 & 1.10 (0.49; 2.46) & 0.810 \\
        Quantile 3 & 0.86 (0.27; 2.74) & 0.798 & 0.97 (0.34; 2.76) & 0.960 \\
        Quantile 4 & 0.44 (0.08; 2.26) & 0.327 & 0.76 (0.16; 3.69) & 0.739 \\
        \midrule
        BCPP & & & & \\
        Quantile 1 & 1.00 & & 1.00 & \\
        Quantile 2 & 2.61 (1.02; 6.68) & 0.051 & 0.60 (0.32; 1.10) & 0.106 \\
        Quantile 3 & 3.48 (0.96; 12.67) & 0.064 & 0.78 (0.34; 1.79) & 0.566 \\
        Quantile 4 & 11.48 (1.38; 95.69) & 0.028 & 1.04 (0.30; 3.69) & 0.945 \\
        \midrule
        BCEP & & & & \\
        Quantile 1 & 1.00 & & 1.00 & \\
        Quantile 2 & 1.22 (0.54; 2.74) & 0.628 & 0.96 (0.49; 1.86) & 0.896 \\
        Quantile 3 & 2.29 (0.83; 6.30) & 0.114 & 1.39 (0.65; 2.97) & 0.401 \\
        Quantile 4 & 2.11 (0.44; 10.14) & 0.354 & 2.10 (0.81; 5.44) & 0.134 \\
        \midrule
        DBuP & & & & \\
        Quantile 1 & 1.00 & & 1.00 & \\
        Quantile 2 & 1.67 (0.68; 4.14) & 0.270 & 1.24 (0.71; 2.18) & 0.453 \\
        Quantile 3 & 2.47 (0.78; 7.85) & 0.132 & 1.02 (0.59; 1.78) & 0.940 \\
        Quantile 4 & 5.89 (1.03; 33.68) & 0.051 & 0.79 (0.34; 1.81) & 0.577 \\
        \bottomrule
        \end{tabular}
        \end{adjustbox}
        \caption*{Age Subgroup Model: Logistic regression adjusted for \texttt{age.cat}, gender, BMI, ethnicity, citizenship, education, marital status, Poverty income ratio, smoking, diabetes, alcohol, activity level, urinary creatinine (log-transformed).}
    \end{table}

    \begin{figure}[h!]
        \centering
        \includegraphics[width=1\linewidth]{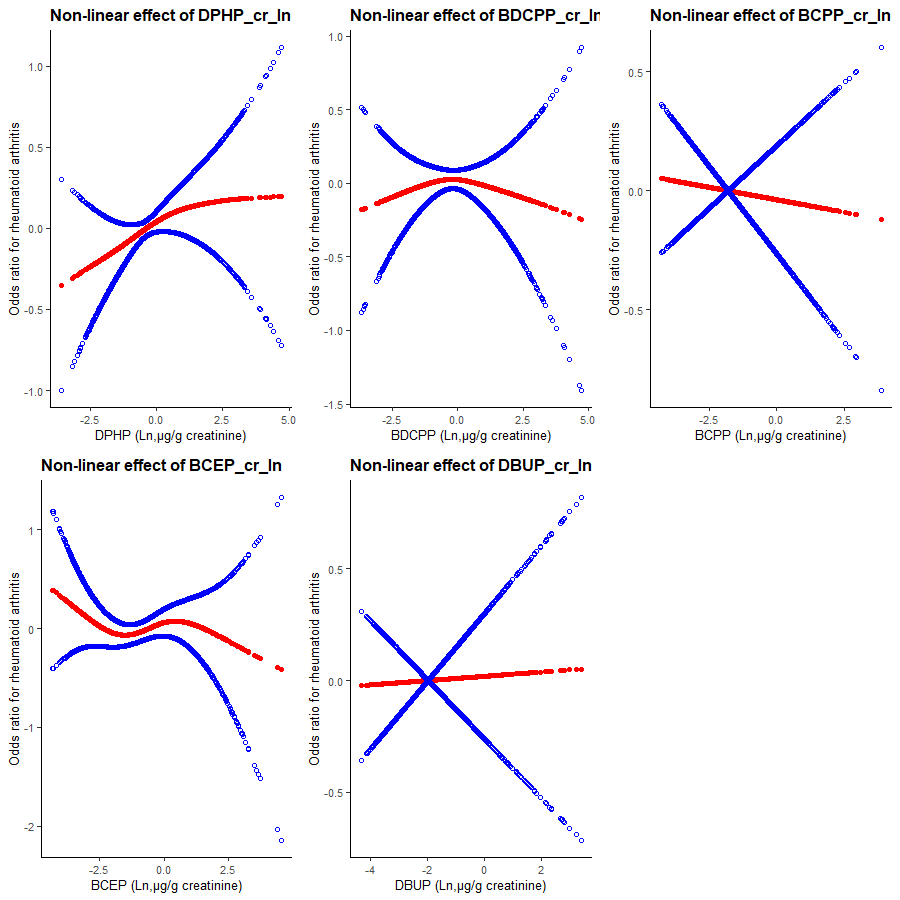}
        \caption{Associations between exposure to OPEs and RA. Blue bands represent the 95\% confidence interval from the fit. The covariates used in the modeling are \texttt{age.cat}, gender, BMI, ethnicity, citizenship, education, marital status, Poverty income ratio, smoking, diabetes, alcohol, activity level, urinary creatinine (log-transformed).}
         \label{fig:nonlinearRelationshipBetweenOPEsMetabolites}
     \end{figure}

\section{Discussion}

    To the best of our knowledge, this is the first large-sample study, a population-based study to investigate the association between exposure to OPEs and rheumatoid arthritis. From the association results, we observed that exposure to OPEs (DPHP and DBUP) was associated with a 1.45 – 1.60-fold increased prevalence of RA, whereas BDCPP appeared to be protective, reducing the risk of RA among participants between 20-60 years. High levels of OPEs were significantly associated with the elevated prevalence in females and participants between the ages of 20-60 years. No significant association was observed between BCPP and BCEP. We also fitted restricted cubic splines and observed a positive correlation between DPHP and DBUP with RA. Rheumatoid arthritis is a chronic autoimmune disease characterized by inflammation, pain, and joint stiffness \cite{elter2020phthalate}. Several studies have suggested a potential association between exposure to organophosphate esters and the development or exacerbation of rheumatoid arthritis. One study conducted on a population of agricultural workers found that those with occupational exposure to OPEs had a significantly higher risk of developing rheumatoid arthritis compared to those without exposure \cite{ostiguy2009chronic}. Another study conducted on a population of Gulf War veterans found a positive correlation between exposure to OPEs during the war and an increased risk of developing rheumatoid arthritis later in life \cite{ronsmans2021associations}. These findings suggest that exposure to OPEs may play a role in the pathogenesis of rheumatoid arthritis \cite{de2009mosaic}. The mechanism by which OPEs may contribute to the development of rheumatoid arthritis is still unclear. Some hypotheses suggest that OPEs may trigger an immune response, producing autoantibodies and subsequent joint inflammation \cite{ospina2018exposure}. However, it is important to note that these studies have limitations, and further research is needed to fully understand the association between OPEs exposure and rheumatoid arthritis. Additionally, there are challenges in studying the relationship between exposure to OPEs and rheumatoid arthritis due to the complex nature of both the disease and the physiochemical activity of OPEs.

    We observed the curves for DPHP and DPHP, which suggest a positive association between higher exposure levels and an increased risk of RA. BCEP had a U-shaped relationship, with an increased risk of RA at lower and higher exposure; however, the associations were insignificant. The curves for BCPP and BDCPP show an initial increase in RA risk with exposure to OPEs and then a decrease at higher levels. The subgroup gender stratification suggests that exposure to OPEs is higher in females than males. The observed association of OPEs and RA herein is biologically plausible as nail polish is considered a significant source of short-term DPHP exposure and chronic exposure for frequent users or those occupationally exposed \cite{mendelsohn2016nail}. Studies have found that exposure to OPEs can disrupt both innate and adaptive immune systems, causing delayed or inappropriate immune responses to pathogens \cite{khani2023cellular}. The proper equilibrium of extracellular immune ligands, which are essential for starting and controlling immune responses, is disrupted by OPEs.

    Epidemiological studies have confirmed that exposure to certain OPEs reduces the adhesion capacity of macrophages, impairing their ability to respond to inflammatory signals and pathogens. According to a study on human cell lines, flame retardant exposure increased the release of pro-inflammatory cytokines like IL-1$\beta$, IL-6, and IL-17 and increased the expression of the membrane protein E-cadherin, which impairs LPS-induced inflammatory responses in macrophages \cite{longo2019vitro}. Alturaiki, in a recent study, found higher levels of rheumatoid factor (RF), C-reactive protein (CRP), and erythrocyte sedimentation rate (ESR) in RA patients \cite{alturaiki2022assessment}. He also found serum levels of IL-1$\beta$, IL-6, IL-8, and CCL5, but not TNF-$\alpha$, significantly increased in those with RA\cite{nishioka2012di}. It is plausible that the production of these disease-relevant pro-inflammatory cytokines is intricately linked to the development of rheumatoid arthritis (RA) \cite{li2018bisphenol}. RA is well-documented for causing progressive destruction of synovial joints, primarily through irreversible damage to articular cartilage and bone. During RA inflammation, cytokines such as IL-1$\beta$, IL-6, and TNF-$\alpha$ play pivotal roles by activating the JAK/STAT pathway, which subsequently increases the expression of receptor activator of nuclear factor kappa-$\beta$ ligand (RANKL) \cite{aletaha2018diagnosis}. The activated JAK/STAT3 pathway also suppresses osteoprotegerin expression, mediated by the Wnt signaling pathway. RANKL, by binding to its receptor RANK on the surface of osteoclast precursors, activates the NF-$\kappa$B pathway, thereby promoting osteoclast differentiation and bone resorption \cite{khani2023cellular}. This cascade of molecular events not only leads to enhanced bone degradation but also perpetuates the inflammatory environment within the joint. Additionally, the chronic inflammation and cytokine milieu in RA can induce synovial fibroblasts to produce matrix metalloproteinases (MMPs), which further degrade the extracellular matrix of the cartilage. This creates a vicious cycle of inflammation and tissue destruction, exacerbating joint damage. Understanding these pathways highlights potential therapeutic targets for RA, such as inhibitors of the JAK/STAT pathway, RANKL-RANK interactions, and MMP activity, aiming to mitigate inflammation and prevent joint destruction \cite{aletaha2018diagnosis, firestein2017immunopathogenesis}. Furthermore, some studies have shown that exposure to flame retardants will potently induce the production of RA-relevant cytokines TNF-$\alpha$, IL-6, and IL-1$\beta$ \cite{li2020vitro}.

\section{Conclusion}

    In conclusion, our results indicate that exposure to flame retardants (DPHP and DBUP) is associated with an increased prevalence of RA, however, inverse association between urinary BDCPP concentration was observed. These associations appear to be more pronounced in female and older people. Given the cross-sectional design of the current study, these results should be interpreted with caution, and further investigations are needed to confirm our findings. 

\appendix

    

\bibliographystyle{elsarticle-num} 
\bibliography{main}





\end{document}